\pdfoutput=1

\documentclass[aps,pra,reprint,showpacs,superscriptaddress]{revtex4-1}
\usepackage{graphicx}
\usepackage{amsmath}
\usepackage{amssymb}
\usepackage{amsfonts}
\usepackage{latexsym}
\usepackage{rotating}
\usepackage{color}
\usepackage{latexsym}
\usepackage{epsfig}
\usepackage{natbib}
\usepackage{bm}
\usepackage{url}

\usepackage{color} 

\usepackage{hyperref}
\hypersetup{
colorlinks=true,final=true,
        linkcolor=blue,
        citecolor=blue,
        filecolor=blue,
        urlcolor=blue,
}

\begin{document}

\title{Perturbed Dissipative Solitons: A Variational Approach}
\author{Ambaresh Sahoo}
\email{ambaresh@phy.iitkgp.ernet.in}
\author{Samudra Roy}
\email{samudra.roy@phy.iitkgp.ernet.in}
\affiliation{Department of Physics, Indian Institute of Technology Kharagpur, W.B. 721302, India}
\author{Govind P. Agrawal}
\affiliation{The Institute of Optics, University of Rochester, Rochester, New York 14627, USA}

\begin{abstract}

We adopt a variational technique to study the dynamics of perturbed dissipative solitons, whose evolution is governed by a Ginzburg--Landau equation (GLE). As a specific example of such solitons, we consider a silicon-based active waveguide in which free carriers are generated through two-photon absorption. In this case, dissipative solitons are perturbed by physical processes such as third-order dispersion, intrapulse Raman scattering, self-steepening, and free-carrier generation. To solve the variational problem, we adopt the Pereira--Stenflo soliton as an ansatz since this soliton is the exact solution of the unperturbed GLE\@. With this ansatz, we derive a set of six coupled differential equations exhibiting the dynamics of various pulse parameters. This set of equations provides considerable physical insight in the complex behavior of perturbed dissipative solitons. Its predictions are found to be in good agreement with direct numerical simulations of the GLE\@. More specifically, the spectral and temporal shifts of the chirped soliton induced by free carriers and intrapulse Raman scattering are predicted quite accurately. We also provide simple analytic expressions of these shifts by making suitable approximations. Our semi-analytic treatment is useful for gaining physical insight into complex soliton-evolution processes.

\end{abstract}

\maketitle

\section{Introduction}

A dissipative soliton is a stable, strongly localized structure forming inside a nonlinear dissipative system under suitable conditions \cite{Akhmediev}. Its applications range from optics, condensed-matter physics, cosmology to biology and medicine. Dissipative solitons arise in an open nonlinear system, far from equilibrium, and a continuous supply of energy is essential for them. More specifically, pulse-like dissipative solitons form inside a nonlinear active medium as a result of double balance between the medium's nonlinearity and its dispersion and between the gain and loss mechanisms that change pulse energy. Owing to this dual balance, the parameters of a dissipative soliton, such as its amplitude, width, chirp, and phase, do not depend on the initial conditions.

Active optical waveguides provide a fertile ground for observing optical dissipative solitons (ODSs) by launching short optical pulses inside them. In practice, however, such ODSs are sensitive to perturbations such as higher-order dispersion and self-steepening that become non-negligible for femtosecond pulses. Another important nonlinear effect for such short pulses is the intrapulse Raman scattering (IRS) that leads to a continuous red-shift of the pulse spectrum. In this paper we study the effects of IRS and other perturbations on the ODS dynamics through a variational approach technique \cite{Bondeson}. The variational technique is a standard method used extensively for both the dissipative \cite{Kaup} and non-dissipative \cite{Anderson} soliton systems. Its application is straightforward for conservative (non-dissipative) systems by choosing a suitable Lagrangian density \cite{Anderson}. The Lagrangian needs to be modified in case of dissipative systems such that it consists of a conservative part and a dissipative part \cite{Cerda}. Construction of a Rayleigh dissipation function is an alternative method to handle the dissipative effects \cite{Royjlt}. In all cases, The Lagrangian density is reduced by integrating over time. This reduction process requires a suitable \textit{ansatz}. The variation technique makes the assumption that the functional form of the ansatz remains intact in presence of a small perturbation but all parameters appearing in the anstaz (amplitude, width, position, phase, frequency etc.) may evolve with propagation. The reduced variational problem, followed by the Ritz optimization, leads to a set of coupled ordinary differential equation (ODE) that governs the evolution of individual pulse parameters under the influence of the perturbation~\cite{GPAbook1}.

A proper choice of the ansatz is critical for success of any variational approach. For example, soliton perturbation theory uses the hyperbolic-secant profile of a Kerr soliton as its ansatz with considerable success~\cite{GPAbook1}. However, this form will not be suitable for ODSs as they represent chirped optical pulses. In this work we adopt the Pereira--Stenflo solution~\cite{PS} of the Ginzburg--Landau equation (GLE) and show that the choice of this solution as an ansatz to the variational problem is much superior compared to the choice of a Kerr soliton. We examine the dynamics of various pulse parameters and predict accurately both the magnitude of the spectral red-shift of the ODS initiated by the IRS and corresponding changes in  its speed. We also show that the ODS undergoes a slight blue-shift when self-steepening acts as a perturbation. The characteristic shift in the ODS location by the third-order dispersion (TOD) is also captured by the variational treatment presented here. As a special case, we consider the ODS formation inside an active silicon waveguide where free carriers are generated through multi-photon absorption and examine the perturbing effects of free carriers on an ODS\@. To verify the accuracy of our variational results, we compare them to the full numerical solution of the GLE and find a reasonable agreement between the two. We also propose some closed-form solutions which may prove more convenient to use in practice.

\section{THEORY}
\label{Ginzburg--Landau Equation}

To be realistic and to take into account several practical perturbations, we choose a silicon-based, active, nano-photonic waveguide \cite{Agazzi} and study the formation and evolution of ODSs in such a system. In such a waveguide. the leading loss mechanism comes from two-photon absorption (TPA) when pumped at a wavelength below 2.2~$\mu$m. As a consequence of TPA, free carriers are generated inside the waveguide that introduce additional loss so-called free-carrier absorption (FCA) and also change the refractive index \cite{Dieter, Tomita} through a phenomenon called free-carrier dispersion (FCD). In our model we take account these effects by coupling the carrier dynamics with the complex GLE that governs the pulse dynamics \cite{Lin,Roy-M-B}. This equation is a kind of nonlinear Schr\"{o}dinger equation with complex coefficients representing growth and damping \citep{PS, Anderson_Scr}. Its classical solution is known as the Pereira--Stenflo soliton \cite{PS} and it constitutes a specific example of dissipative solitons.

The extended GLE describing evolution of optical pulses inside a silicon-based active waveguide can be written in the following normalized form \cite{Roy-M-B, GPAbook2},
\begin{align} \label{gl}
    i\frac{\partial u}{\partial \xi }-\frac{1}{2}sgn\left( {{\beta }_{2}} \right)\frac{{{\partial }^{2}}u}{\partial {{\tau }^{2}}}-i\left( {{g}_{0}}+{{g}_{2}}\frac{{{\partial }^{2}}}{\partial {{\tau }^{2}}} \right)u+i\alpha u \nonumber \\
    +\left( 1+iK \right){{\left| u \right|}^{2}}u -i{{\delta }_{3}}\frac{{{\partial }^{3}}u}{\partial {{\tau }^{3}}} -\tau_R u \frac{\partial {\left| u \right|}^{2}}{\partial \tau } \nonumber \\
    +i s\frac{\partial{({\left| u \right|}^{2} u)}}{\partial \tau} +\left( \frac{i}{2}-\mu  \right){{\phi }_{c}}u=0
\end{align}
where the free-carrier effects are included through the normalized density parameter $\phi_c$ that satisfies the rate equation \cite{Lin},
\begin{equation} \label{ansatz}
    d\phi_c/d\tau=\theta|u|^4-\tau_c \phi_c  .
\end{equation}
The time and distance variables are normalized as $\tau=t/t_0$ and $\xi=z/L_D$, where $t_0$ is the initial pulse width and $L_D=t_0^2/|\beta_2 (\omega_0)|$ is the dispersion length, $\beta_2(\omega_0)$ being the group-velocity dispersion coefficient at the carrier frequency  $\omega_0$.

The preceding equations contain multiple dimensionless parameters. The TOD, IRS and self-stepping parameters are normalized as $\delta_3=\beta_3/(3!|\beta_2| t_0)$, $\tau_R=T_R/t_0$ and $s=1/(\omega_0 t_0)$, where $T_R$ is the first moment of the Raman response function \cite{GPAbook1}. The field amplitude ($A$) is rescaled as, $A = u\sqrt{P_0}$, where peak power, $P_0=|\beta_2 (\omega_0 )|/(t_0^2 \gamma_R)$, $\gamma_R=k_0 n_2/A_{eff}$ and $n_2\approx(4\pm 1.5)\times 10^{-18} \ m^2 W^{-1}$ is the Kerr-nonlinear coefficient of silicon. The dimensionless TPA coefficient is given as,  $K=\gamma_I/\gamma_R=\beta_{TPA}\lambda_0/(4\pi n_2)$, where, $\beta_{TPA}\approx 8\times 10^{-12} \ m W^{-1}$ and $\gamma_I=\beta_{TPA}/(2A_{eff})$.  The linear loss coefficient is normalized as $\alpha=\alpha_l L_D$. The free-carrier density $N_c$ is related to $\phi_c$ as $\phi_c=\sigma N_c L_D$ where $\sigma\approx 1.45\times 10^{-21}\ m^2$ is the FCA cross section of silicon at $\lambda_0=1.55~\mu m$ \cite{Rong}. The generation of free carriers is regulated by the parameter  $\theta=\beta_{TPA} |\beta_2 |\sigma/(2\hbar\omega_0 A_{eff}^2 t_0 \gamma_R^2)$ \cite{Lin-Z-P}. The parameter  $\mu=2\pi k_c/(\sigma \lambda_0)$ is the FCD coefficient with $k_c\approx 1.35\times 10^{-27}\ m^3$ \cite{Dinu}. The carrier recombination time $t_c$ is scaled as $\tau_c=t_0/t_c$. The gain $G$ and the gain dispersion coefficient ($g_2$) are normalized as  $g = GL_D$ and $g_2= g(T_2/t_0)^2$, where dephasing time is $T_2$. The spectral wings of the pulse experience less gain due to a finite gain bandwidth related to $g_2$.

In the absence of TOD ($\delta_3=0$), IRS ($\tau_R=0$), self-steepening ($s=0$) and free carriers (i.e., $\phi_c=0$), Eq.\ (1) reduces to the standard GLE, which is known to have the stable ODS solution in the following form \cite{PS, GPAbook2, Desurvire}:
\begin{equation} \label{ansatz}
    u\left( \xi, \tau \right)={{u}_{0}}{{\left[ \text{sech}\left( \eta \tau  \right) \right]}^{\left( 1+ia \right)}}{{e}^{i\text{ }\!\!\Gamma\!\!\text{ }\xi }},
\end{equation}
where the four parameters $u_0,\ \eta,\ a$ and $\Gamma$ are given by~\cite{GPAbook2}:
\begin{subequations} \label{ansatz_part}
\begin{align}
|u_0|^{2}& = \frac{(g_{0}-\alpha)}{K}\left[ 1-\frac{sgn(\beta_{2})a/2 + g_{2}} {{g}_{2} ({{a}^{2}}-1 )- sgn(\beta_2)a} \right], \\
\eta^{2} &= \frac{({{g}_{0}}-\alpha )}{ \left[{{g}_{2}}\left( {{a}^{2}}-1 \right)-sgn\left( {{\beta }_{2}} \right)a \right]}, \\
\Gamma &= \frac{{{\eta }^{2}}}{2}\left[ sgn\left( {{\beta }_{2}} \right)\left( {{a}^{2}}-1 \right)+4a{{g}_{2}} \right], \\
a &= \frac{H-\sqrt{{{H}^{2}}+2{{\delta }^{2}}}}{\delta}.
\end{align}
\end{subequations}
Here, $H=-[(3/2)sgn(\beta_2 )+3g_2 K]$ and $\delta=-[2g_2-sgn(\beta_2)K]$. The preceding solution was first obtained in 1977 and is known as the Pereira--Stenflo soliton~\cite{PS}

\section{VARIATIONAL ANALYSIS}
\label{perturbative}

The ODS solution exists only when four terms in Eq.\ \eqref{gl} related to TOD ($\delta_3=0$), IRS ($\tau_R=0$), self-steepening ($s=0$) and free carriers ($\phi_c=0$) are neglected. The important question is how these terms affect the ODS solution. One can study their impact by solving Eq.\ \eqref{gl} numerically. However, this approach hinders any physical insight. In this section we treat the four terms as small perturbations and study their impact through a variational analysis. The Variational method has been used with success in the past for many pulse-propagation problems \cite{Bondeson,Kaup,Anderson,Cerda,Royjlt}. It requires a suitable \textit{ansatz} for the pulse shape and makes the assumption that the functional form of the pulse shape remains intact in presence of small perturbations but its parameters appearing in the ansatz (amplitude, width, position, phase, frequency etc.) may evolve with propagation. For our problem, it is natural that we choose the Pereira--Stenflo solution in Eq.\ \eqref{ansatz}  as our ansatz since it is the exact solution of Eq.\ \eqref{gl} in the absence of perturbations induced by TOD, IRS, self-steepening ($s=0$) and free-carrier generation. We thus choose the following ansatz:
\begin{align}\label{nls0}
 u\left( \xi ,~\tau  \right)={{u}_{0}}\left( \xi  \right){{\left[ \text{sech}\left\{ \eta \left( \xi  \right)\left( \tau -{{\tau }_{p}}\left( \xi  \right) \right) \right\} \right]}^{\left\{ 1+ia\left( \xi  \right) \right\}}} \nonumber \\
exp\left[ i\left\{ \phi \left( \xi  \right)-\text{ }\Omega\left( \xi  \right)\left( \tau -{{\tau }_{p}}\left( \xi  \right) \right) \right\}  \right],
\end{align}
where the six parameters $u_0,~\eta,~\tau_p,~\phi,~a$ and $\Omega$ are now assumed to depend on $\xi$. We first write Eq.~\eqref{gl} in the form of a perturbed nonlinear Schr\"{o}dinger equation~\cite{Anderson, GPAbook1}:
\begin{equation} \label{nls1}
    i\frac{\partial u}{\partial\xi}+\frac{1}{2}\frac{\partial^{2}u}
    {\partial\tau^{2}}+ |u|^{2}u = i\epsilon(u),
\end{equation}
where we have chosen the dispersion to be anomalous $(\beta_2<0)$ and define $\epsilon(u)$ as,
\begin{align} \label{nls2}
    \epsilon(u) = \delta_3\frac{\partial^3u}{\partial{{\tau}^{3}}}-i\tau_R & \frac{\partial{|u|}^{2}}{\partial\tau} -s\frac{\partial ({|u |}^{2} u)}{\partial \tau} -\left(\frac{1}{2}+i\mu\right)\phi_c u \nonumber \\
   &~~~~~ -K|u|^{2}u +{g}_{0}u +{{g}_{2}}\frac{{{\partial }^{2}}}{\partial {{\tau }^{2}}}u.
\end{align}
We then follow a standard procedure \cite{GPAbook1} and introduce the Lagrangian density appropriate for Eq.\ \eqref{nls1} and integrate over $\tau$ using the ansatz in Eq.\ \eqref{nls0} to obtain the following reduced Lagrangian:
\begin{gather}
    L = \frac{2{{u}_{0}}^{2}}{\eta }\left( \frac{\partial \phi }{\partial \xi }+\Omega \frac{\partial {{\tau }_{p}}}{\partial \xi } \right)-\frac{a {{u}_{0}}^{2}}{{{\eta }^{2}}}\frac{\partial \eta }{\partial \xi }+C\frac{{{u}_{0}}^{2}}{\eta }\frac{\partial a}{\partial \xi } \nonumber \\
    +\frac{\eta {{u}_{0}}^{2}}{3}\left( 1+{{a}^{2}} \right)+\frac{{{u}_{0}}^{2}}{\eta }\left( {{\Omega }^{2}}-\frac{2}{3}{{u}_{0}}^{2} \right) \nonumber \\
    +i\int_{-\infty }^{\infty}(\epsilon{u}^{*}-\epsilon^* u)\,d\tau,
\end{gather}
where $C=[\ln(4)-2]$. The next step is to use the Euler-Lagrange equation for each pulse parameter to obtain a set of coupled ODEs for the six parameters that describe the overall soliton dynamics \cite{GPAbook1,Hasegawa-K}. These equations govern the evolution of pulse energy $(E=\int_{-\infty}^{\infty} |u|^{2}\,d\tau)$, temporal position $\tau_p$, frequency shift $\Omega$, amplitude $\eta$, frequency chirp $a$, and phase $\phi$ and have the form
\begin{align}
    \frac{dE}{d\xi }&=\frac{d}{d\xi }\left( \frac{2{{u}_{0}}^{2}}{\eta} \right) =2{\rm Re}\int\limits_{-\infty }^{\infty }{\epsilon {{u}^{*}}}\,d\tau, \label{var1}\\
    \frac{d{{\tau}_{p}}}{d\xi }& = -\Omega +\frac{\eta }{{{u}_{0}}^{2}} \int\limits_{-\infty }^{\infty }{\left( \tau -{{\tau }_{p}} \right){\rm Re}\left( \epsilon {{u}^{*}} \right)\,d\tau }, \label{var2}\\
    \frac{d\Omega }{d\xi} &= \frac{{{\eta }^{2}}}{{{u}_{0}}^{2}}\int\limits_{-\infty }^{\infty }{\tanh\left[ \eta \left( \tau -{{\tau }_{p}} \right) \right]{\rm Re}\left[ \left( a+i \right)\epsilon {{u}^{*}} \right]\,d\tau }, \label{var3}\\
    \frac{d\eta }{d\xi} &= \frac{2{{\eta }^{2}}}{{{u}_{0}}^{2}}\int\limits_{-\infty }^{\infty }{\ln\left[ {\rm sech}\left\{ \eta \left( \tau -{{\tau }_{p}} \right) \right\} \right]{\rm Re}\left( \epsilon {{u}^{*}} \right)\,d\tau } \nonumber \\
&\hspace{.5cm}  -C\frac{\eta^{2}}{2u_{0}^{2}}E_\xi+\frac{2{{\eta}^{3}}a}{3}, \label{var4}
\end{align}
\begin{align}
\frac{da}{d\xi}& = -\frac{a\eta }{2{{u}_{0}}^{2}}E_\xi +\frac{2}{3}{{u}_{0}}^{2}-\frac{2}{3}{{\eta }^{2}}\left( 1+{{a}^{2}} \right) +\frac{\eta }{{{u}_{0}}^{2}}{\rm Im}\int\limits_{-\infty }^{\infty }{\epsilon {{u}^{*}}}\,d\tau \nonumber \\
  &\hspace{0cm} -\frac{2{{\eta }^{2}}}{{{u}_{0}}^{2}}{\rm Im}\int\limits_{-\infty }^{\infty }{\left( \tau -{{\tau }_{p}} \right)tanh\left[ \eta \left( \tau -{{\tau }_{p}} \right) \right]\left( 1-ia \right)\epsilon {{u}^{*}}\,d\tau }, \label{var5}\\
\frac{d\phi}{d\xi} &= \frac{a}{2\eta}\eta_\xi -\frac{C}{2}a_\xi  -\Omega \,{\tau_p}_\xi + \frac{2}{3}u_0^2 -\frac{1}{6}\eta^2(1+a^2) -\frac{1}{2}\Omega^2 \nonumber \\ &\hspace{0.5cm}  +\frac{\eta}{2u_0^2}{\rm Im}\int\limits_{-\infty }^{\infty}{\epsilon {{u}^{*}}}\,d\tau, \label{var6}
\end{align}
where Re and Im stand for real and imaginary parts. The final step is to evaluate all the integrals using $\epsilon(u)$ given in Eq.\ \eqref{nls2}. It results in the following set of six coupled differential equations:
\begin{align}
\frac{dE}{d\xi }&=\frac{2}{3}(3g_0-K\eta E)E- \frac{2}{3}g_2[(1+a^2)\eta^2+3\Omega^2]E  \nonumber \\&\hspace{0.5cm} -\frac{1}{6}\theta \eta E^3 , \label{var7} \\
\frac{d{{\tau }_{p}}}{d\xi }&=-\left(1 +2{{g}_{2}}a \right)\Omega -\frac{7}{72}\theta E^2 +\delta_3\left[(1+a^2)\eta^2 + 3\Omega^2 \right]  \nonumber \\&\hspace{0.5cm} +\frac{1}{2}s \eta E , \label{var8}\\
\frac{d\Omega }{d\xi }&=-\frac{4}{3}{{g}_{2}}\left( 1+{{a}^{2}} \right)\Omega {{\eta }^{2}}   -\frac{4}{15}\tau_R E{{\eta }^{3}} + \frac{4}{15}s a E \eta^3 \nonumber \\&\hspace{0.5cm} + \frac{2}{15}\left(\mu -\frac{a}{2}\right)\theta \eta^2 E^2 , \label{var9}\\
\frac{d\eta }{d\xi }&=\frac{2}{3} (a-EK)\eta^2-\frac{4}{9}(2-a^2)g_2\eta^3 -\frac{1}{6}C\theta \eta^2 E^2 \nonumber \\ &\hspace{0.5cm} -4\delta_3 a \Omega \eta^3,\label{var10}\\
\frac{da}{d\xi }&=\frac{1}{3}(1+aK)E\eta -\frac{2}{3} (1+ag_2)(1+a^2)\eta^2 -\frac{1}{6}\theta a \eta E^2 \nonumber \\ &\hspace{0.5cm} + \frac{1}{3}s \Omega \eta E +4\delta_3 \Omega \eta^2(1+ a^2), \label{var11}\\
\frac{d\phi}{d\xi}&=\frac{a}{2\eta}\eta_\xi -\frac{C}{2}a_\xi  -\Omega{\tau_p}_\xi + \frac{1}{3}\eta E -\frac{1}{6}\eta^2(1+a^2)-\frac{1}{2}\Omega^2 \nonumber \\ &\hspace{-0.2cm} -\frac{1}{6}\mu \theta \eta E^2 +\frac{1}{3}s\Omega E \eta + \delta_3 \left[(1+a^2)\Omega \eta^2 + \Omega^3 \right].
\label{var12}
\end{align}

These equations provide considerable physical insight since they show which perturbations affect a specific pulse parameter. For example, the Raman parameter $\tau_R$ appears only in the equation for the frequency shift $\Omega$ and the term containing it has a negative sign. This immediately shows that the IRS leads to a spectral red-shift of the ODS\@. In contrast, the self-steepening parameter $s$ appears in the frequency equation in a term with the positive sign and shows that self-steepening will reduce the Raman-induced spectral red-shift. This kind of physical insight is very valuable in interpreting the numerical results. It is noteworthy that the phase $\phi$ does not appear in any equation except the last one. This indicates that the numerical value of the soliton's phase does not affect any of its other parameters. For this reason, we ignore the phase equation in the following discussion. In the next section we discuss the effects of various perturbations on the evolution of the ODS parameters and also compare variational results with the results obtained from direct simulation of Eq.\ \eqref{gl}.

\section{Full Numerical Simulations}

\begin{figure*}[tb!]
\begin{center}
\epsfig{file=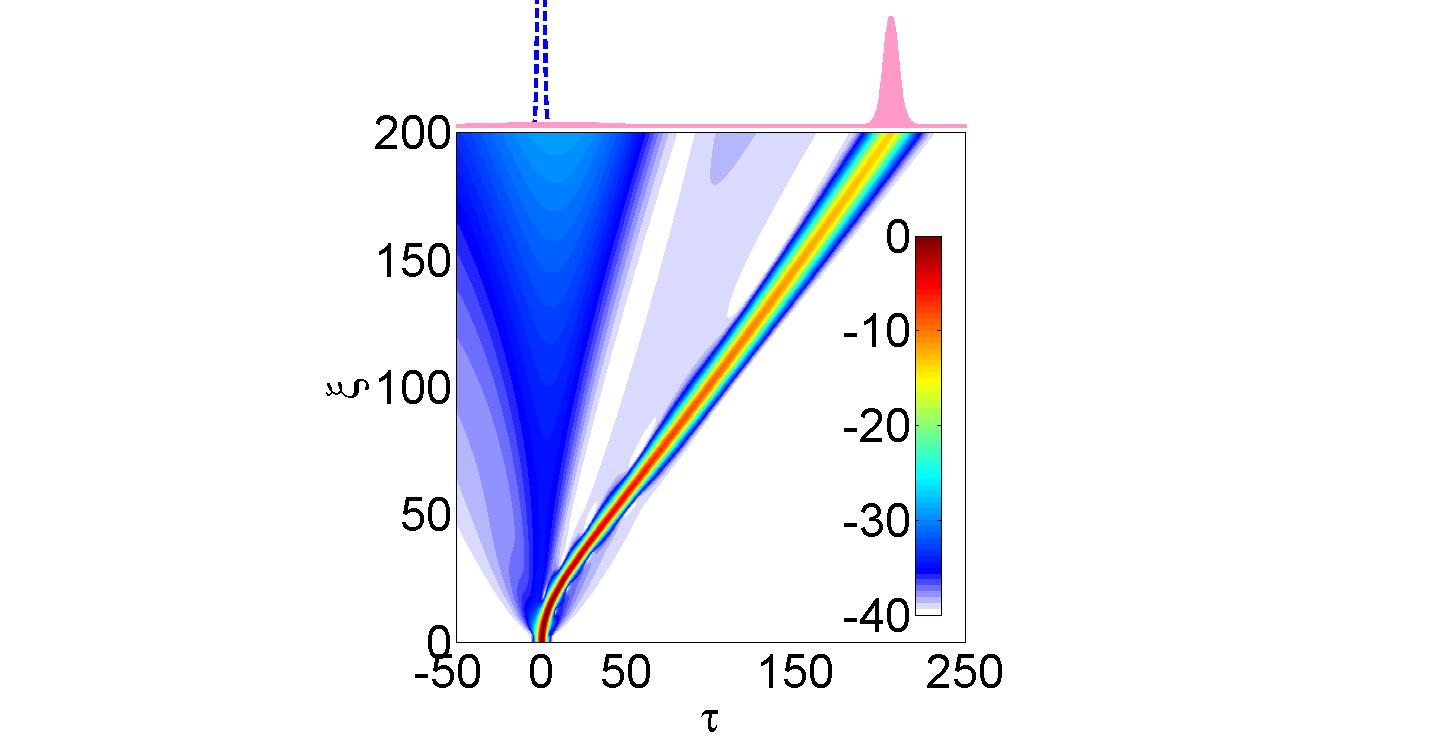,trim=4.5in 0.05in 6.05in 0.0in,clip=true, width=59mm}
\epsfig{file=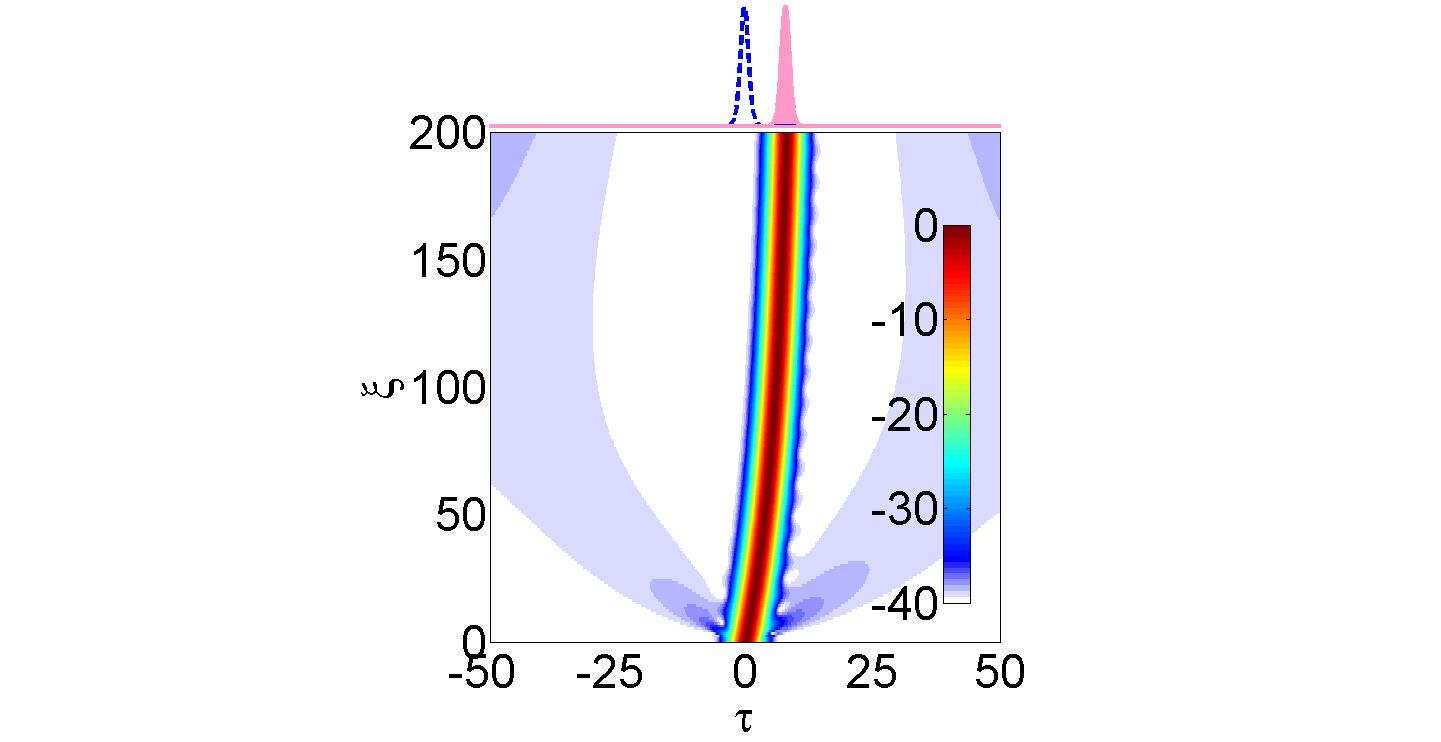,trim=5in 0.05in 5.7in 0.0in,clip=true, width=58mm}
\epsfig{file=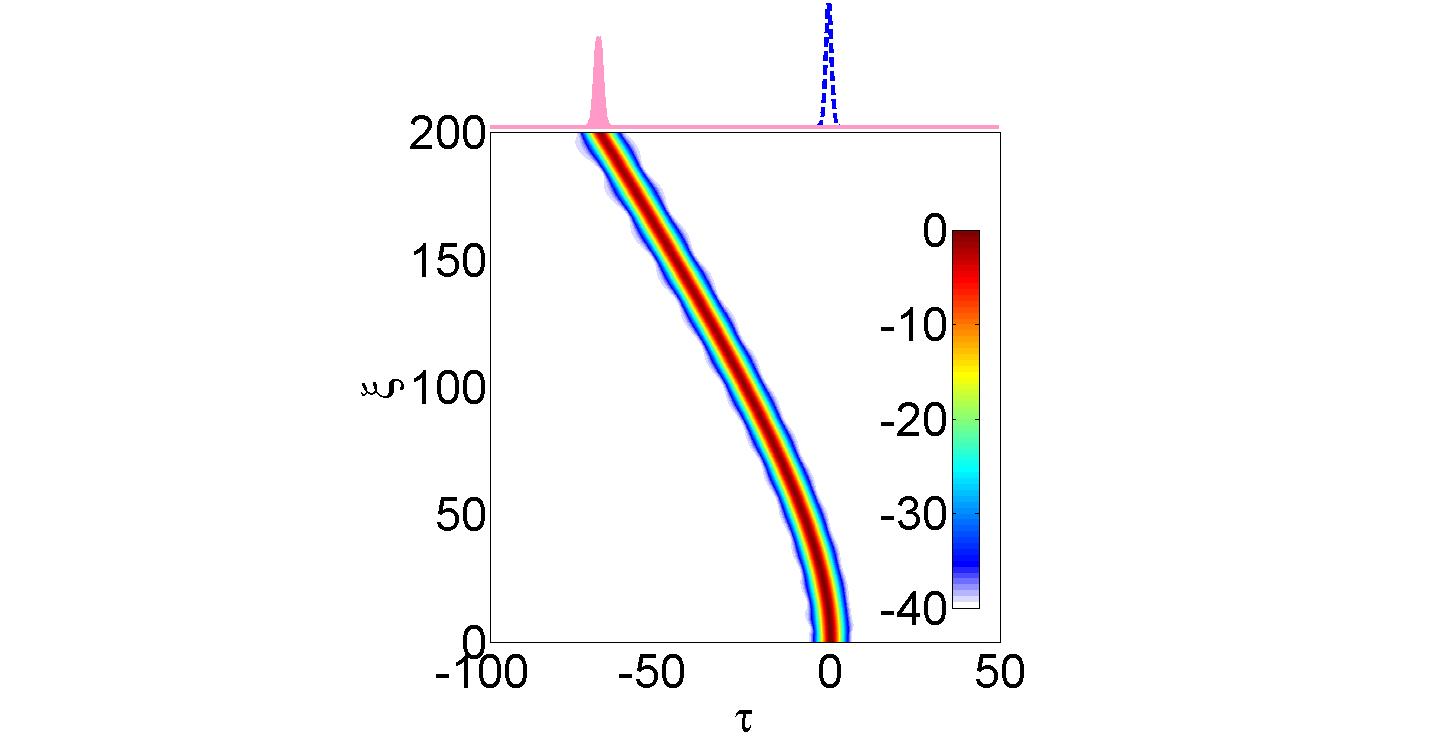,trim=5in 0.05in 5.7in 0.0in,clip=true, width=58mm}
\vspace{0.5em}
~~~~~~~~(a)~~~~~~~~~~~~~~~~~~~~~~~~~~~~~~~~~~~~~~~~~~~~~~~~~~~(b)~~~~~~~~~~~~~~~~~~~~~~~~~~~~~~~~~~~~~~~~~~~~~~~~~~~~~(c)
\\
\epsfig{file=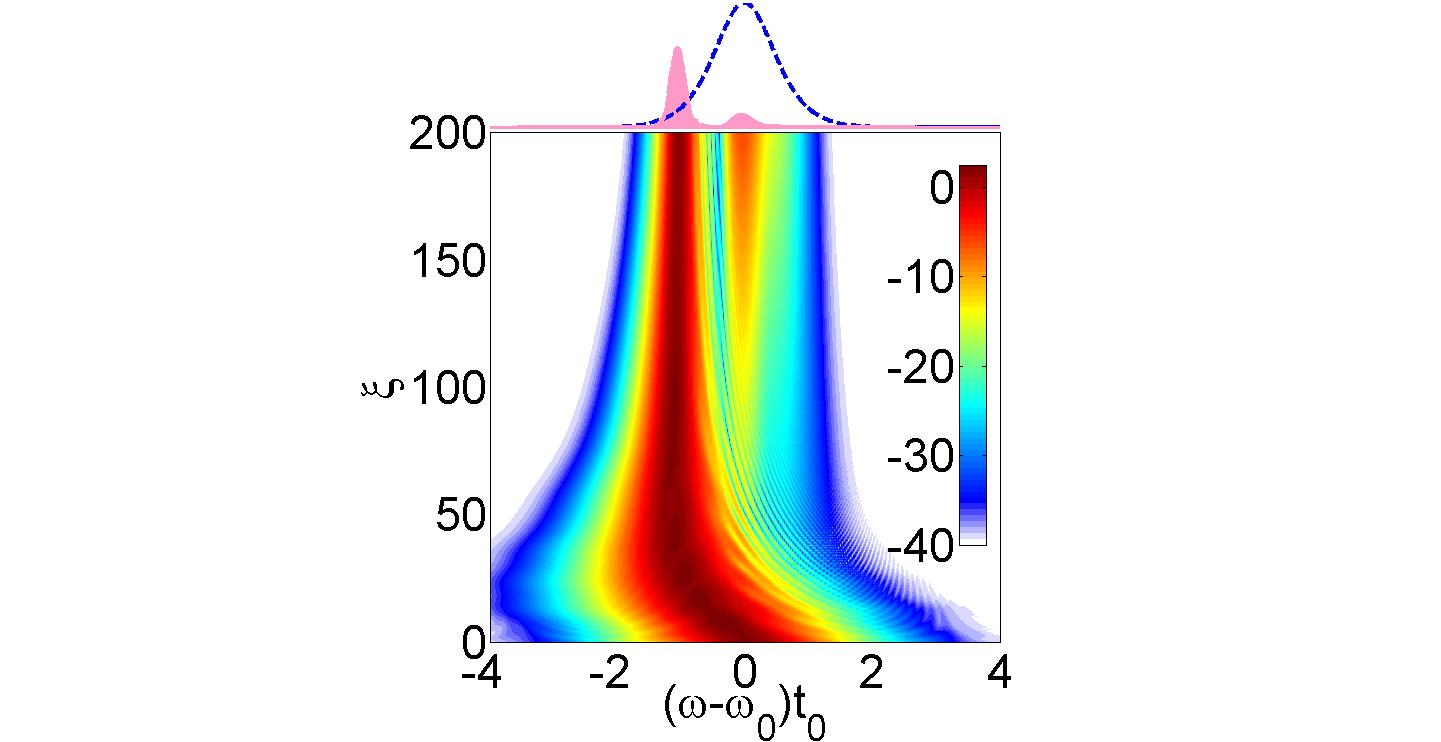,trim=5in 0.00in 5.7in 0.0in,clip=true, width=59mm}
\epsfig{file=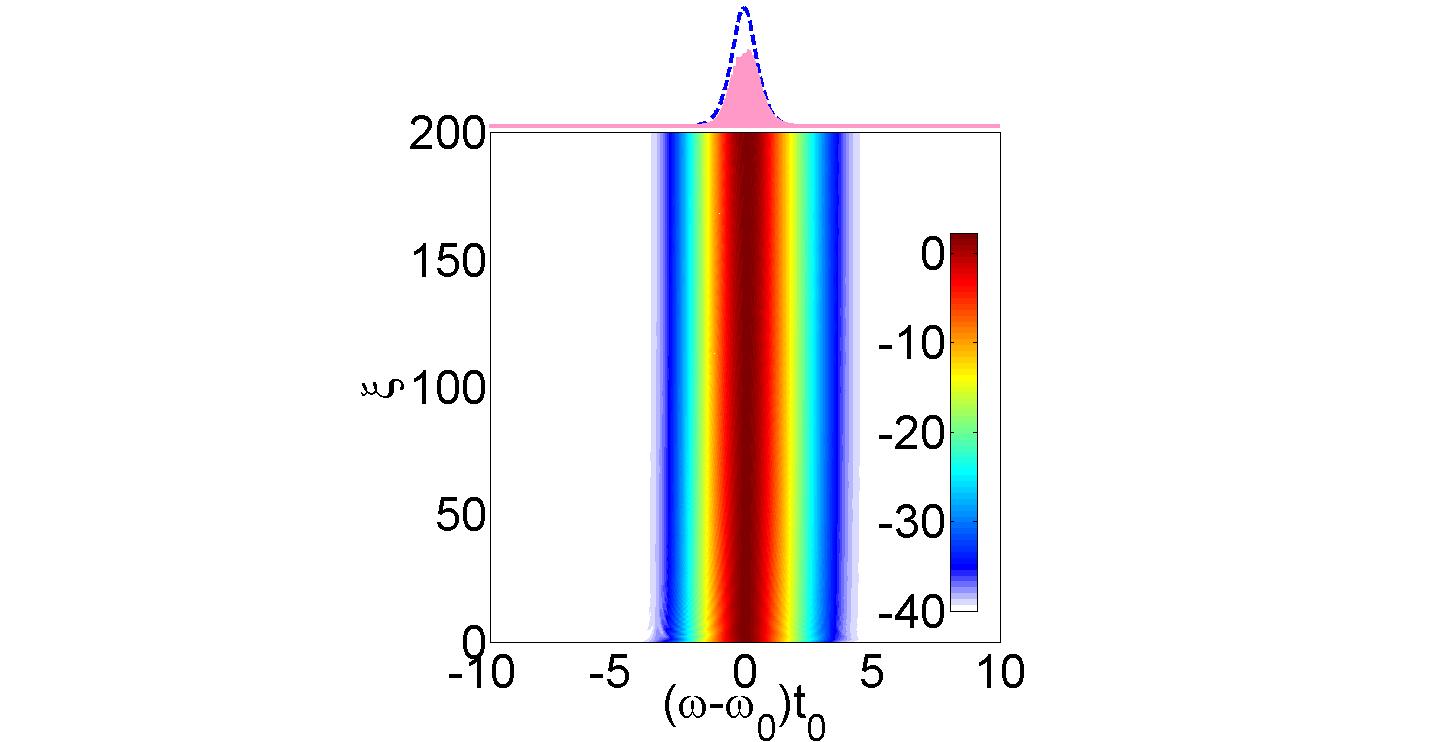,trim=5in 0.00in 5.7in 0.0in,clip=true, width=59mm}
\epsfig{file=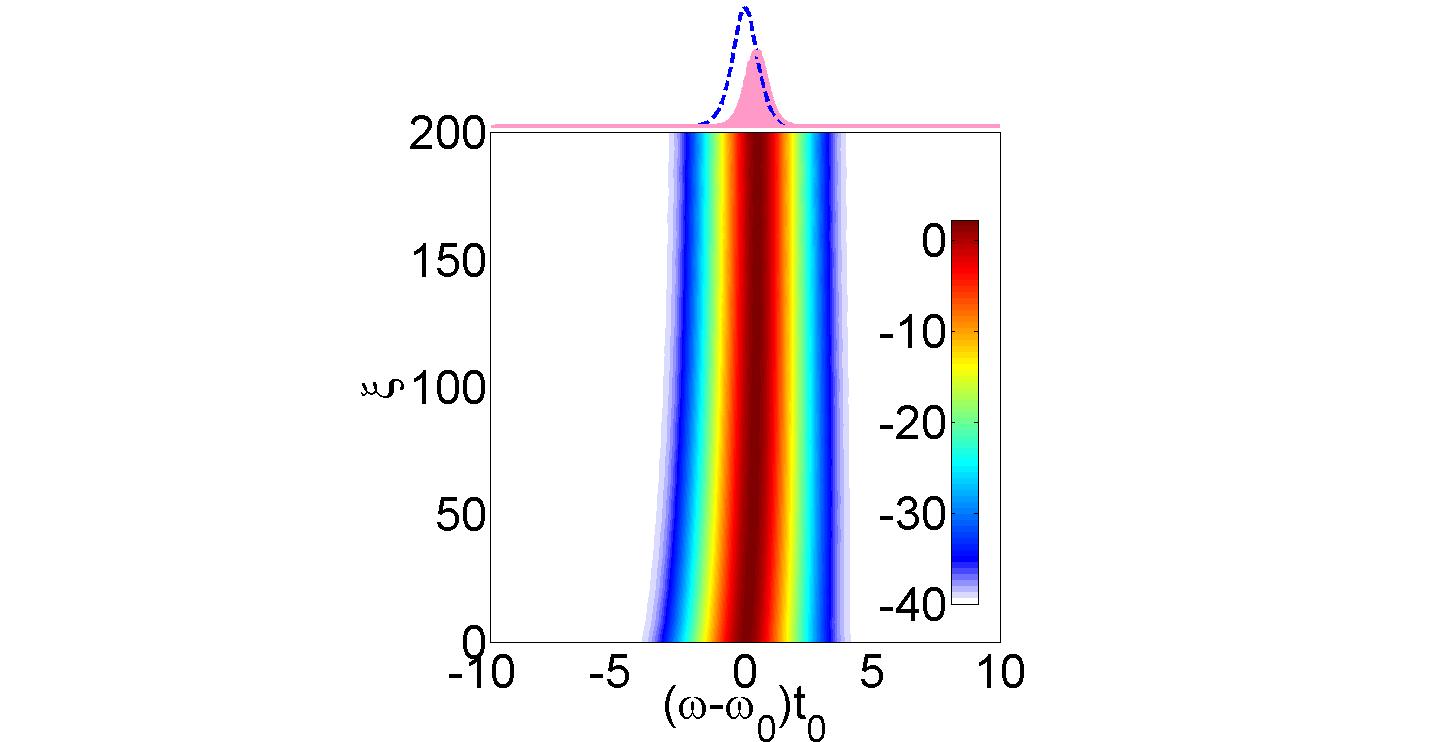,trim=5in 0.00in 5.7in 0.0in,clip=true, width=59mm}
\vspace{0.5em}
~~~~~~~~(d)~~~~~~~~~~~~~~~~~~~~~~~~~~~~~~~~~~~~~~~~~~~~~~~~~~~(e)~~~~~~~~~~~~~~~~~~~~~~~~~~~~~~~~~~~~~~~~~~~~~~~~~~~~~(f)
\caption{(Color online) Temporal (top) and spectral (bottom) evolution of an ODS in three cases under a single perturbation. (a, d) IRS acts alone with $\tau_R=0.1$; (b, e) self-steepening acts alone with $s=0.1$; (c, f) FCD acts alone with  $\theta= 0.0044$. Other parameters used in the simulations are: $K=0.01,~g_0=0.01,~g_2=0.01$ and $\alpha=0$. The input (dotted trace) and output pulse shapes are also shown in the top panel.}\label{variational_all}
\end{center}
\end{figure*}

Before discussing the variational results, we solve Eq.~\eqref{gl} numerically and present the results for a realistic silicon active waveguide. More specifically, the individual and collective effects of various perturbation on the evolution of an ODS are discussed in this section. Since the temporal shape of the ODS is distorted rapidly in presence of TOD, which violets the basic assumption behind the variational technique, initially we study pulse dynamics by setting $\delta_3=0$ in Eq.~\eqref{gl}. We solve this equation with the standard split-step Fourier method \cite{GPAbook1} by taking the input pulse in the form of a Pereira--Stenflo soliton with the parameters given in Eq.\ (\ref{ansatz_part}). The parameter values used were  $K=0.01,~g_0=0.01$, $g_2=0.01$, $\tau_R=0.1$, $s=0.1$, $\theta=0.0044$, and $\mu=3.7741$. The values of $\theta$ and $\mu$ are calculated by adopting the realistic values of device parameters.

Figure~\ref{variational_all} shows the temporal (top row) and spectral (bottom row) evolutions of the perturbed ODS in three cases: (a, d) only IRS, (b, e) only self-steepening, and (c, f) only free carriers perturb the ODS\@. As expected, IRS leads to a spectral red-shift and slows down the ODS considerably. However the red-shift saturates after some distance of propagation (around $\xi=20$). In the time domain, the ODS continues to shift because of a change in its speed induced by the red-shift. Our simulations confirm that the pulse width is also affected by the Raman term. In the case of self-steepening, the shape of the pulse remains almost intact and ODS slows down a bit even though its spectrum undergoes a small blue-shift. The influence of free carriers is more dramatic because of FCD that leads to a larger blue-shift with an acceleration of the pulse, consistent with the previously reported results~\cite{Roy-M-B}.

\section{Results of Variational Analysis}

\begin{figure*}[t!]
\begin{center}
\epsfig{file=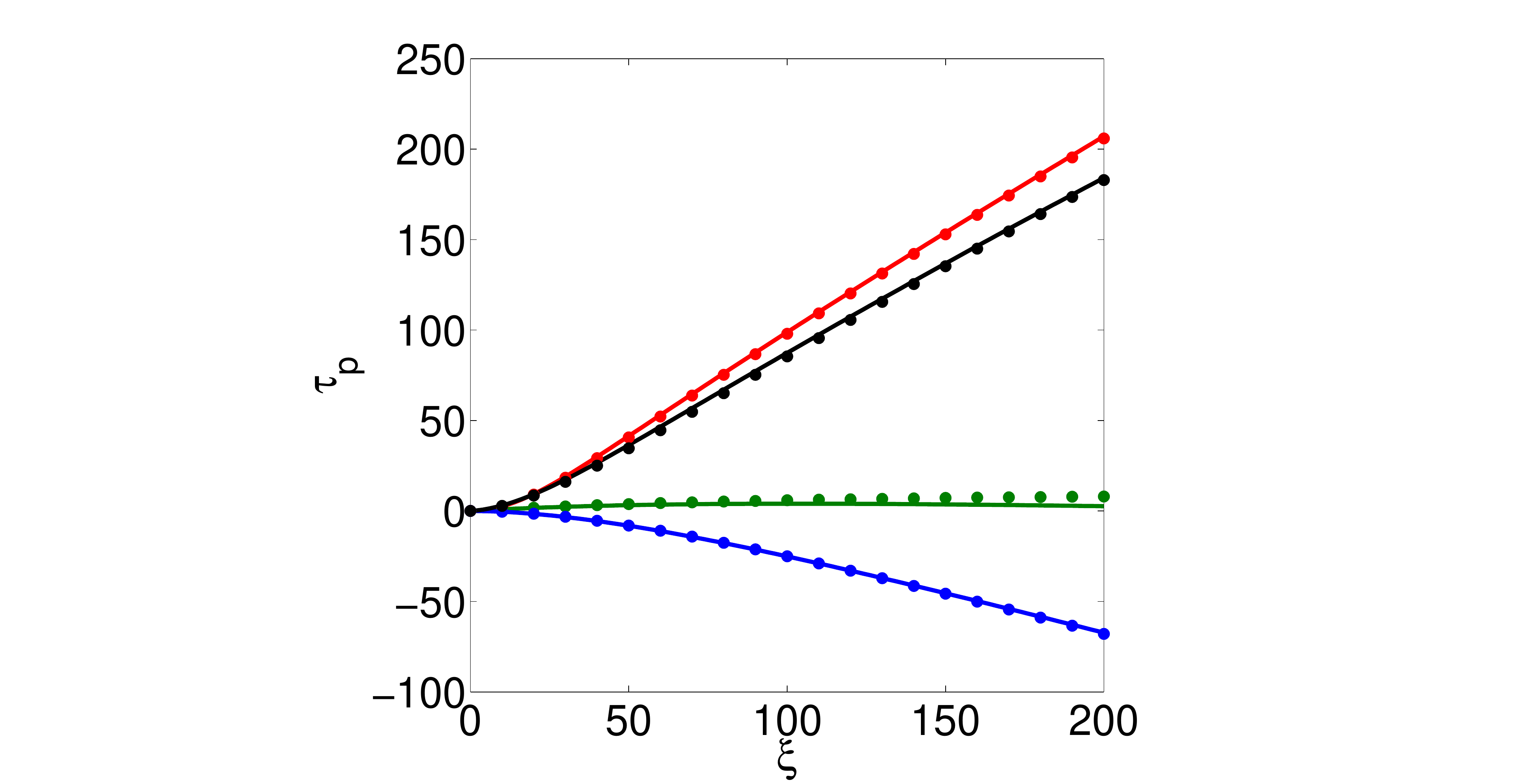,trim=3.0in 0in 3.7in 0in,clip=true, width=60mm}
\epsfig{file=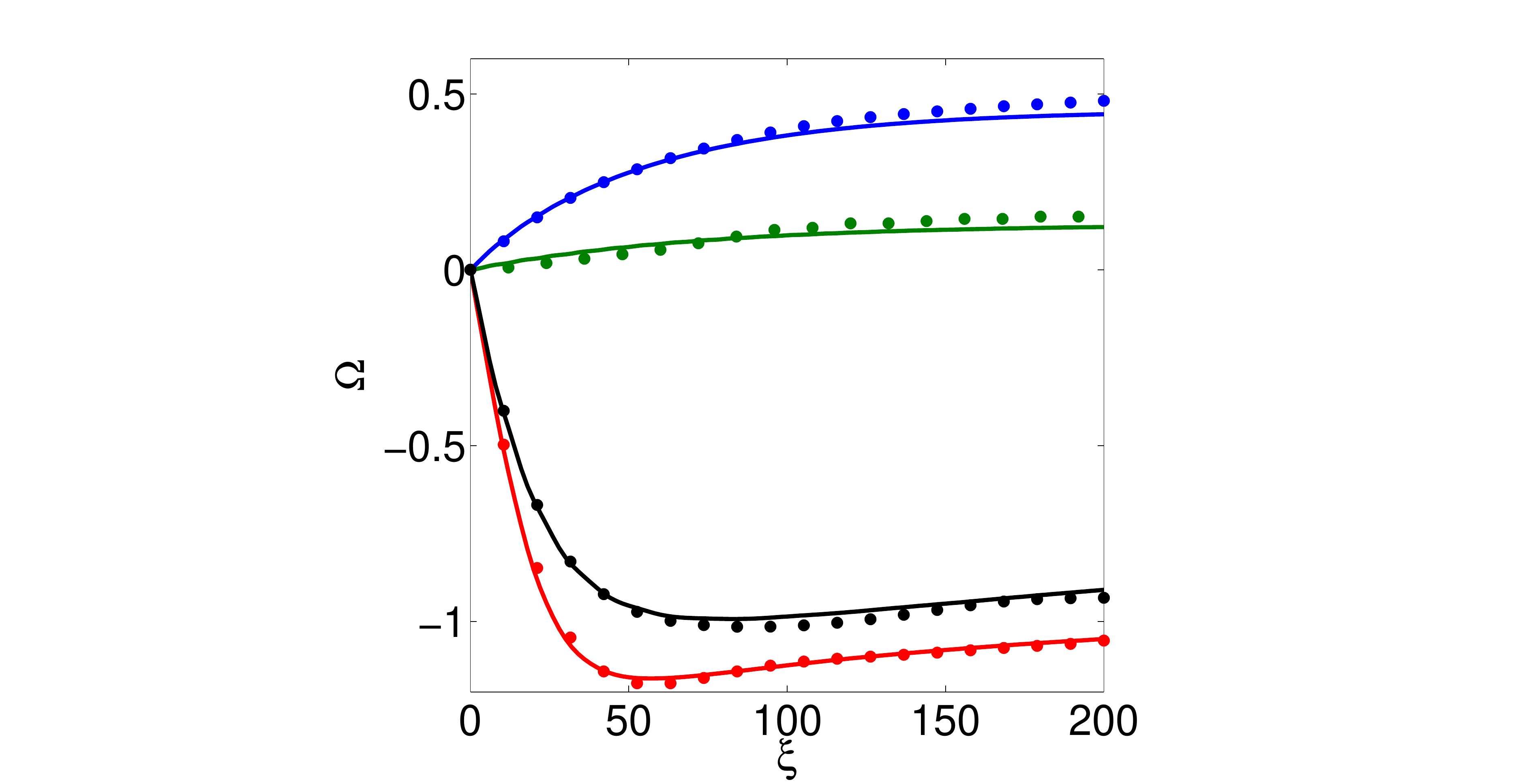,trim=3.0in 0in 3.7in 0in,clip=true, width=60mm}
\vspace{0.5em}
~~~~~~~~~(a)~~~~~~~~~~~~~~~~~~~~~~~~~~~~~~~~~~~~~~~~~~~~~~~~~~~~(b)
\\
\epsfig{file=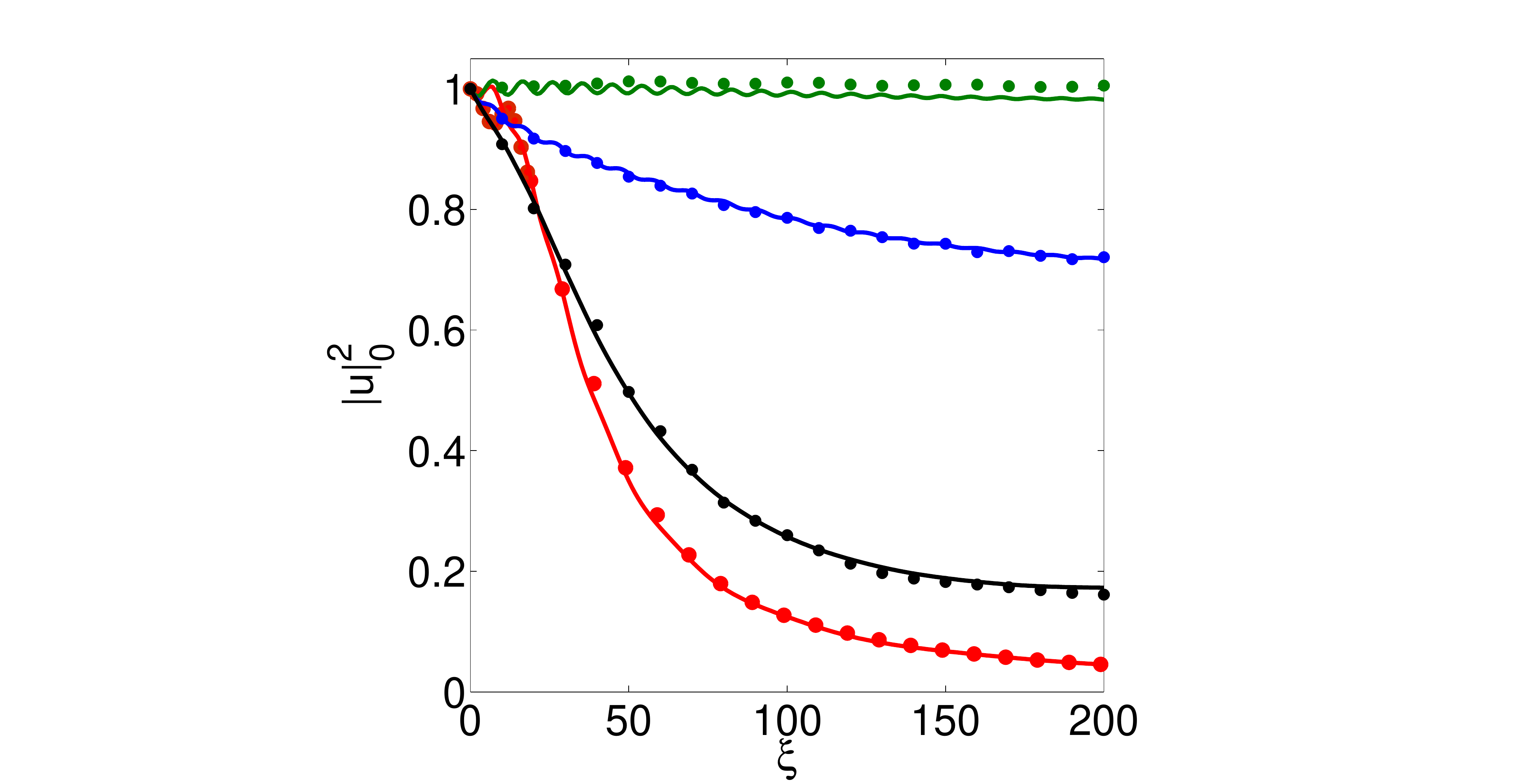,trim=3.0in 0in 3.7in 0in,clip=true, width=60mm}
\epsfig{file=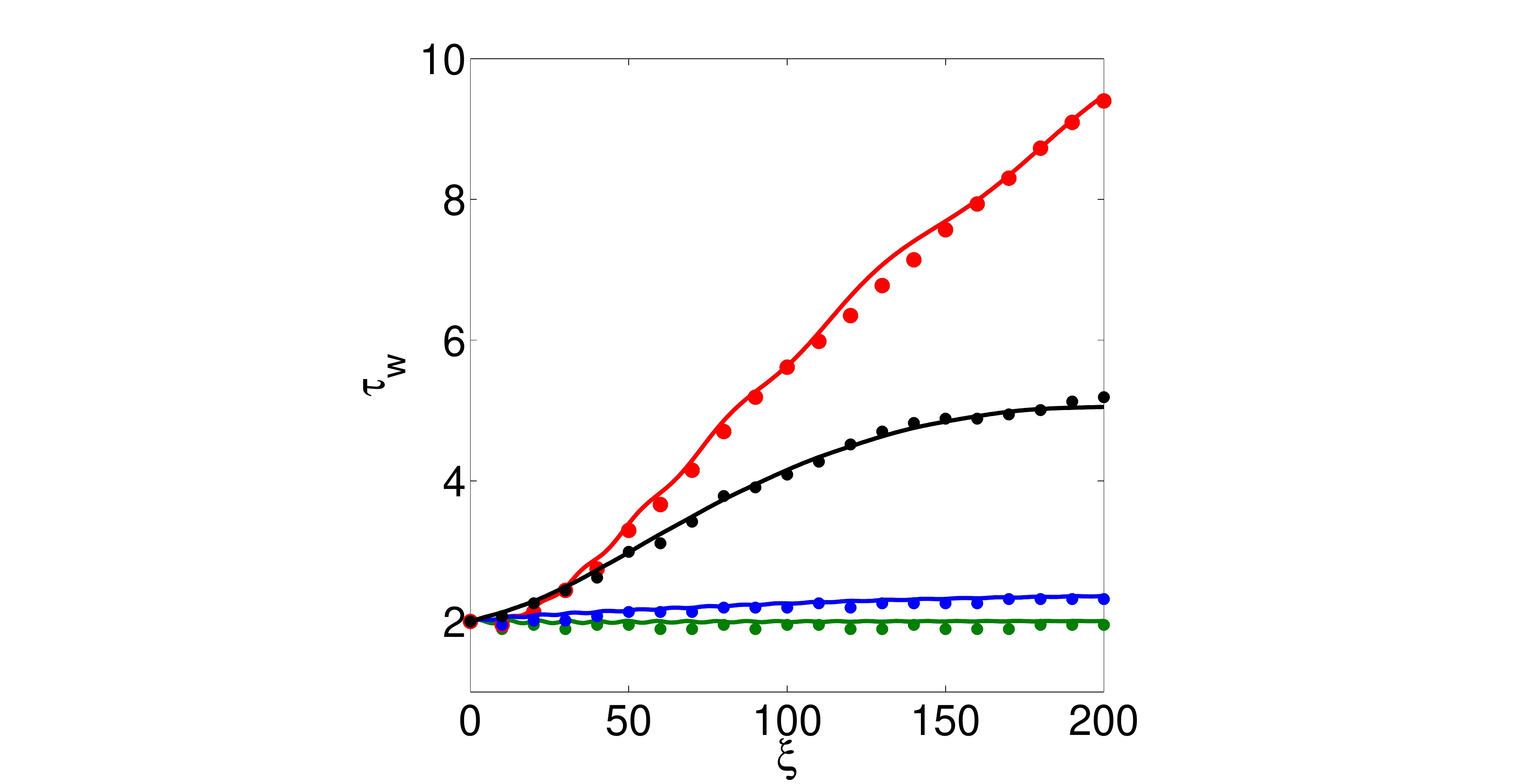,trim=3.0in 0in 3.7in 0in,clip=true, width=60mm}
\vspace{0.5em}
~~~~~~~~~(c)~~~~~~~~~~~~~~~~~~~~~~~~~~~~~~~~~~~~~~~~~~~~~~~~~~~~(d)
\\
\epsfig{file=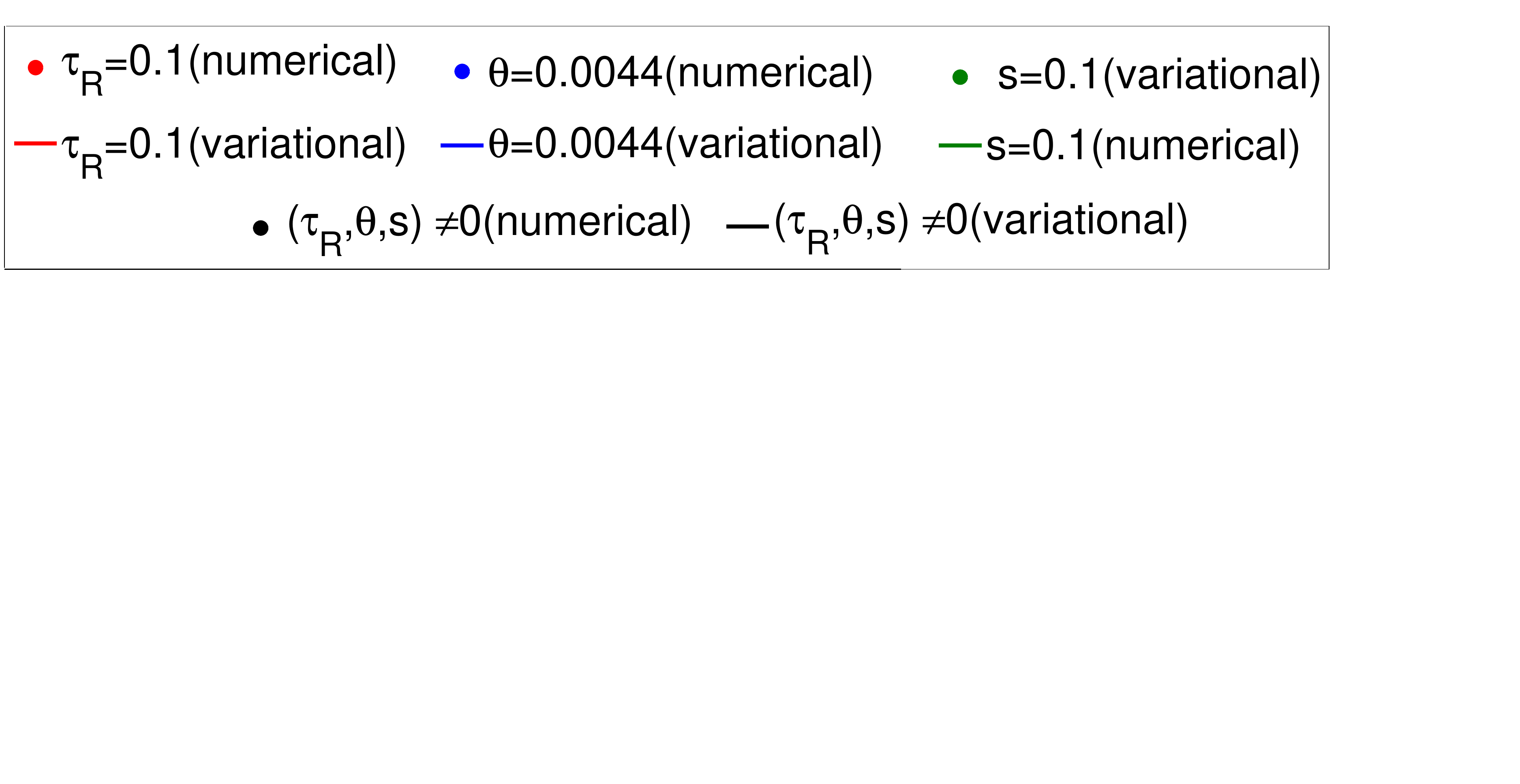,trim=-0.9in 5in 1.7in 0in,clip=true, width=110mm}
\vspace{0.5em}
\caption{(Color online) (a) Temporal position ($\tau_p$) and (b) frequency shift ($\Omega$) as a function of distance in four different cases listed at bottom. Changes in the peak intensity ($|u|^2_0$) and pulse width ($\tau_w$) in the same four cases are shown in parts (c) and (d) respectively. Solid lines show the variational predictions whereas circles represent the corresponding numerical data. Other parameters used in the simulations are: $K=0.01,~g_0=0.01,~g_2=0.01$ and $\alpha=0$.}
\label{variational_mm}
\end{center}
\end{figure*}

In this section we solve the coupled differential equations obtained with the variational approach [Eq.~\eqref{var7}-\eqref{var12}] and compare their predictions with the numerical simulations in Fig.~\ref{variational_mm}. The four parts of this figure compare changes in the pulse position $\tau_p$, spectral shift $\Omega$, peak intensity $|u|^2_0$, and the pulse width $\tau_w$. The red, blue, and green curves in each case correspond to the three cases shown in Fig.\ \ref{variational_all} when only physical process perturbs the ODS\@. The black curves show the case when all three perturbations are present simultaneously. In all cases, the solid lines show variational result and solid circles show the numerical predictions of Eq.~\eqref{gl}. The agreement between the variational and numerical results is remarkably good under so many diverse situations, indicating the suitability of our variational approach for perturbed ODSs.

\begin{figure}[tb!]
\begin{center}
\epsfig{file=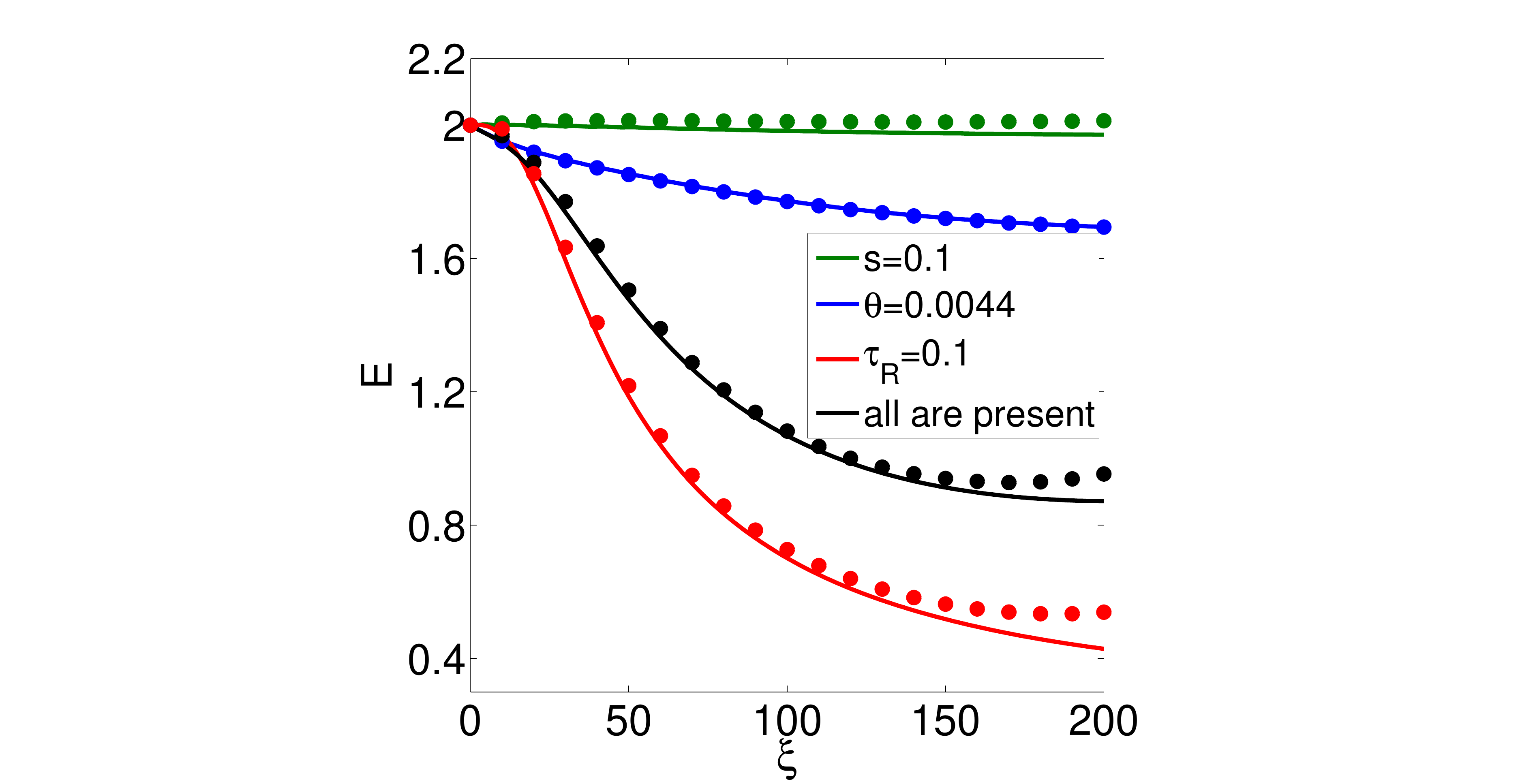,trim=3.0in 0in 3.7in 0in,clip=true, width=44mm}\epsfig{file=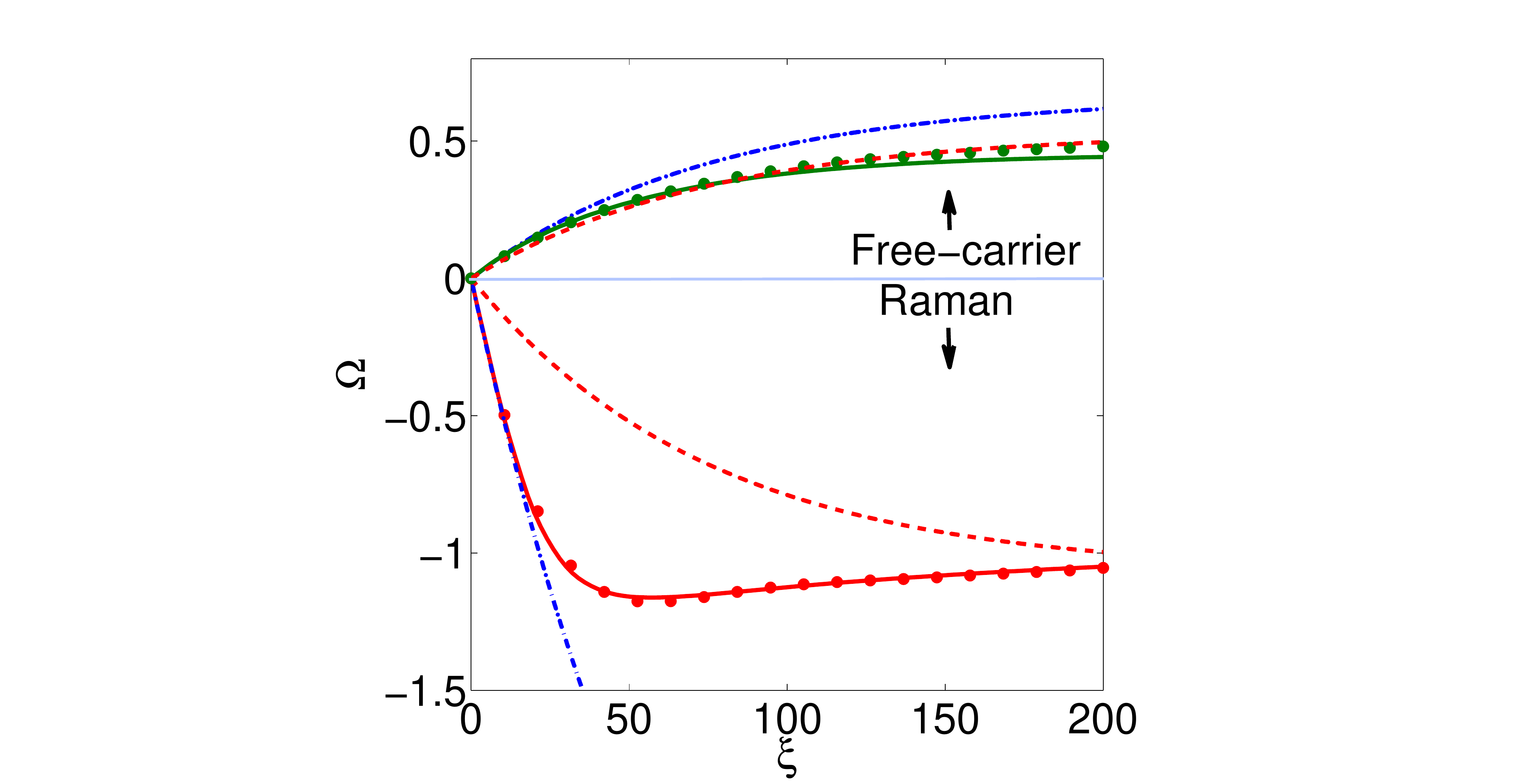,trim=3.0in 0in 3.7in 0in,clip=true, width=44mm}
\vspace{0.5em}
~~~~~~~(a)~~~~~~~~~~~~~~~~~~~~~~~~~~~~~~~~~~~~(b)
\caption{(Color online) (a) Changes in pulse energy as a function of  $\xi$ in the four cases of Fig.~\ref{variational_mm}. (b) Changes in the IRS-induced red shift and FCD-induced blue shift as predicted by the approximated analytic expressions.}
\label{closed_form}
\end{center}
\end{figure}

The set of coupled differential equations becomes more useful if we decouple them with suitable approximations. If we assume variations of $\eta$ and $a$ are relatively small (which is true for propagation distance $\xi<40$) and treat them as constants, we can integrate Eq.~\eqref{var9} analytically. The spectral red-shift owing to IRS can then be written in a close form as,
\begin{equation} \label{nls4}
  \Omega(\xi)\approx -\Omega_R(1-e^{-\rho \xi}),
\end{equation}
where $\Omega_R=\tau_R E_{av}\eta/[5g_2(1+a^2)]$ and $\quad \rho=4g_2(1+a^2)\eta^2/3$. This equation shows how the red-shift increases with $\xi$ initially but saturates to a final value of $-\Omega_R$ when $\xi$ is large enough that $\rho\xi\gg 1$. Here changes in the total energy $E$ are approximated by its average over the distance at which $\Omega$ is calculated. Under the same assumptions, we can integrate Eq.~\eqref{var8} for the temporal shift analytically to obtain
\begin{equation} \label{nls5}
    \tau_p(\xi) \approx \Omega_R(1+2g_2a)[\xi-\rho^{-1}(1-e^{-\rho\xi})].
\end{equation}
This equation shows that once the red-shift satuarates ($\rho\xi\gg 1$),  $\tau_p$ varies linearly with $\xi$; this is clearly evident in Fig.~\ref{variational_mm}(a).

In the same way we can derive an approximate analytic expression for the spectral blue-shift induced by FCD\@. The results has the same form as for IRS, and the blue-shift is,
\begin{equation} \label{nls6}
    \Omega(\xi)\approx\Omega_{FC}(1-e^{-\rho\xi}),
\end{equation}
where the saturated value becomes $\Omega_{FC}=(\mu-a/2)\theta E_{av }^2/[10g_2(1+a^2)]$. The temporal shift due to the FC can also be approximated as,
\begin{equation} \label{nls7}
    \tau_p(\xi)\approx -\Omega_{FC}(1+2g_2a)[(1+\chi_{FC})\xi-\rho^{-1}(1-e^{-\rho\xi})].
\end{equation}

Where $\chi_{FC}=35g_2(1+a^2)/[36(\mu-a/2)(1+2g_2a)]$. The preceding results use the concept of average pulse energy to account for energy variations inside the waveguide in an average sense. In Fig.~\ref{closed_form}(a) we plot energy variations under different perturbations for the 4 cases shown in Fig.~\ref{variational_mm}. Depending on the distance and the mechanism involved, pulse energy may be reduced by more than 50\%. In part (b) we plot the spectral shifts under when IRS and free carriers act as perturbations and compare the full numerical results with the approximate analytical expressions derived above. The agreement with simulations is reasonable in the case of FCD when we use the average energy (red dashed curves). If we use the initial value of pulse energy, agreement is good at short distances but becomes increasingly poor for longer distances (blue dashed curve). In the Raman case, the red dashed curve disagrees initially with numerical results but merges asymptotically to the saturated value predicted by the full calculation. The mismatch  at the initial stage occurs because we assumed $\eta$ to be constant, which is not the case. We emphasize that our closed form expressions help us to understand the pulse dynamics qualitatively. However, the inclusion of all variations of $a$, $\eta$ and $E$ is essential for accurate results.

\begin{figure}[tb!]
\begin{center}
\epsfig{file=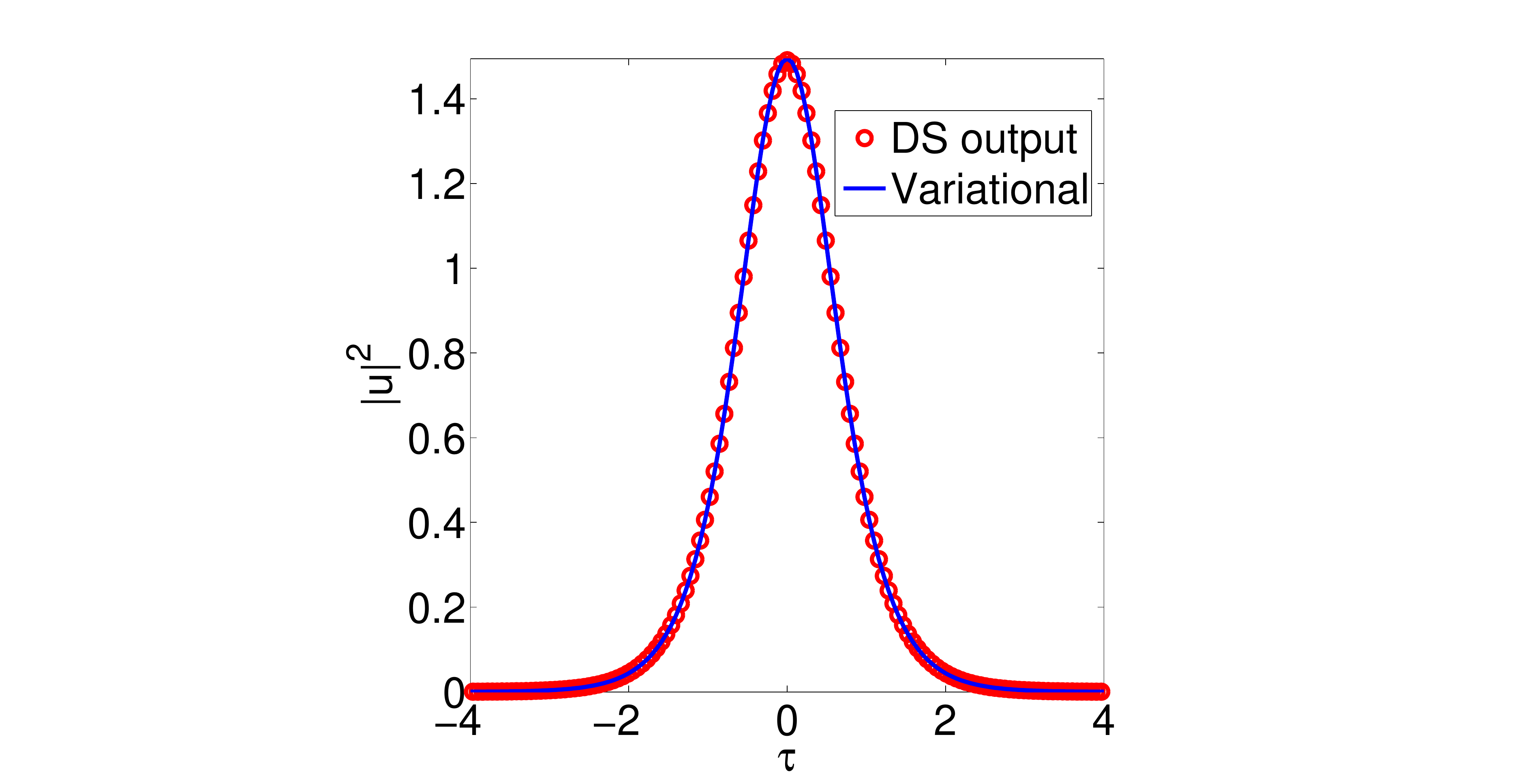,trim=3.0in 0in 3.7in 0in,clip=true, width=44mm}\epsfig{file=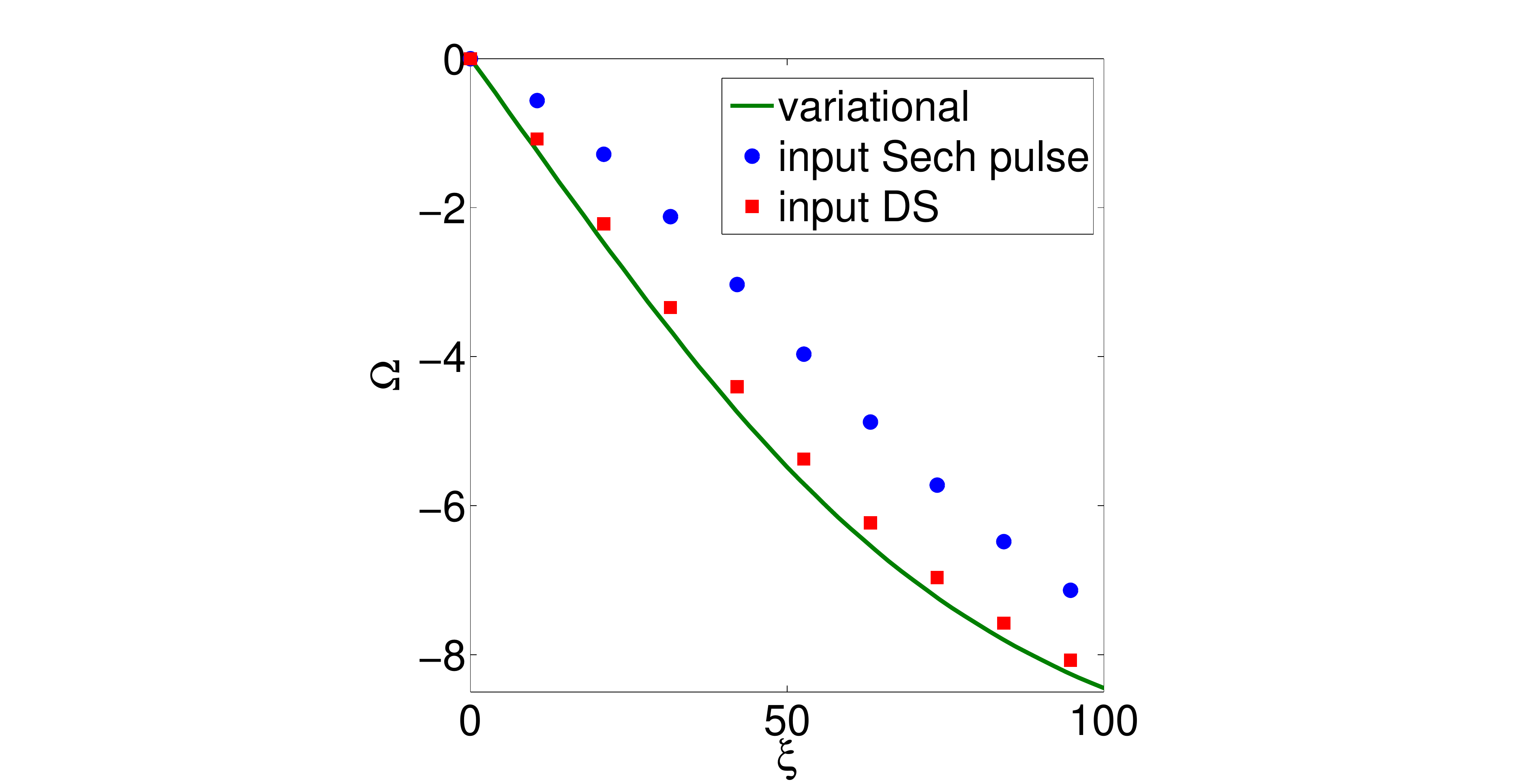,trim=3.0in 0in 3.7in 0in,clip=true, width=44mm}
\vspace{0.5em}
~~~~~~~(a)~~~~~~~~~~~~~~~~~~~~~~~~~~~~~~~~~~~~(b)
\caption{(Color online) (a) Comparison of output intensity profile at $\xi=200$ between full numerical simulation (red dots) and variational prediction (blue trace). (b) Frequency shift as a function of distance for $\tau_R=0.1$. Blue circles and red squares correspond to sech-shape pulse and ODS inputs respectively. The solid green line shows variational prediction of the frequency shift. Parameters used in the simulations are: $K=0.01,~g_0=0.01,~g_2=0.0001$ and $\alpha=0$.}
\label{fig_further}
\end{center}
\end{figure}

As a final test of the set of ODEs derived variationally, we solve them under zero perturbation. If the derived ODEs are correct, they should provide the exact Pereira--Stenflo soliton when all perturbations are switched off. Figure \ref{fig_further}(a) shows that this is indeed the case. The simulated temporal profile at $\xi=200$ overlaps exactly with the variational temporal profile when there is no perturbation. We stress that the use of ODS at the input end is essential while solving Eq.\ (1). In Fig.~\ref{fig_further}(b) we compare the IRS-induced red-shifts obtained using a standard soliton and the ODS at the input end. It is evident that the variational results are consistent with the data obtained using the Pereira--Stenflo soliton as an input but not when a sech-profile of a standard soliton is used for solving Eq.\ (1).

\section{Impact of TOD}

So far, we have ignored the TOD perturbations. However, our variational analysis includes the TOD effects through the $\delta_3$ parameter. Indeed, the temporal position $\tau_p$ of the ODS depends explicitly on $\delta_3$ in Eq.~\eqref{var8}. If we ignore all other perturbations and set $\Omega=\theta=s=0$ in this equation, we get a simple relation $\tau_p(\xi) \approx \delta_3(1+a^2)\eta^2\xi$, provided both $\eta$ and $a$ remain nearly constant. It shows that TOD shifts the soliton position linearly with distance, a well-known result for the standard solitons. To see if this linear behavior persists for an ODS, we solve Eq.\ (1) by taking TOD as the only perturbation ($\tau_R=s=\theta=0$).

\begin{figure}[tb!]
\begin{center}
\epsfig{file=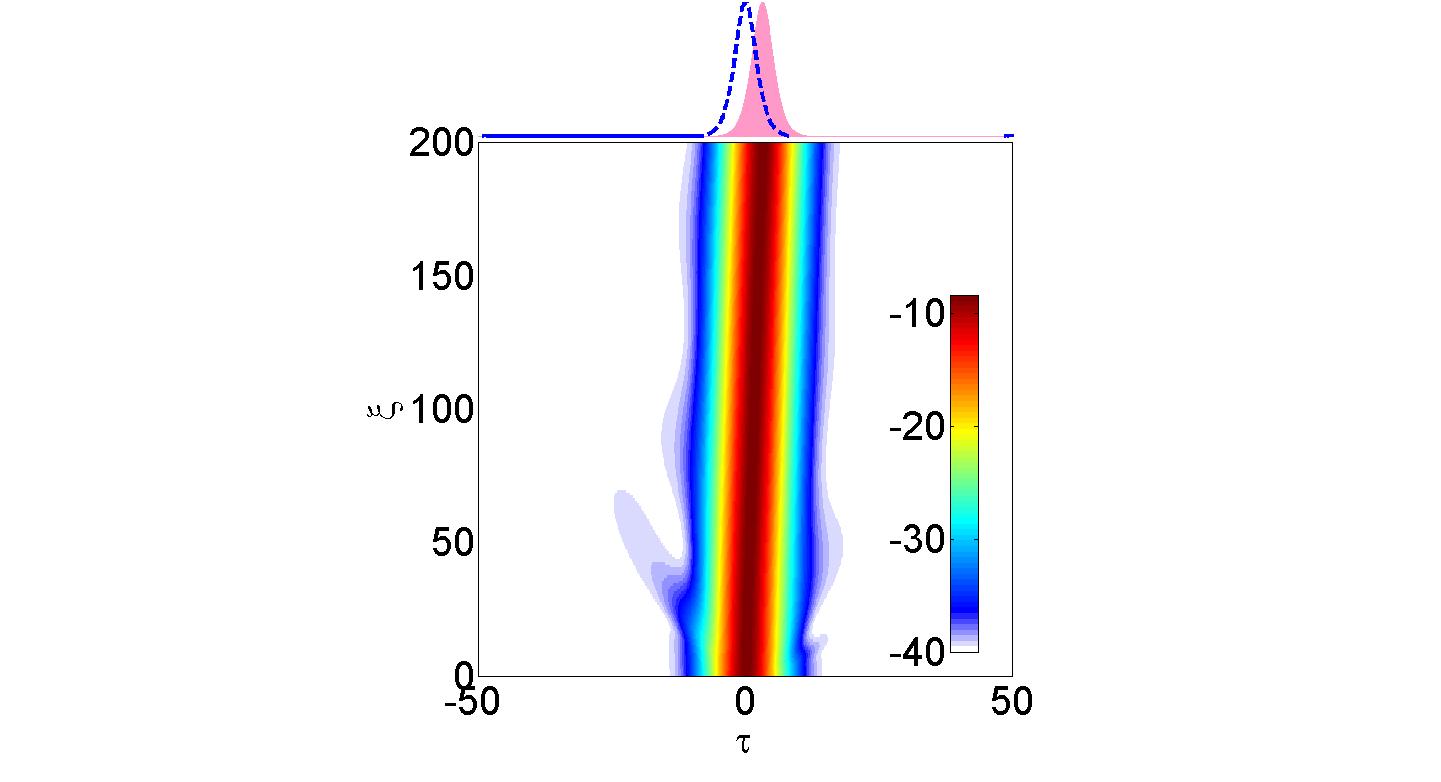,trim=5.1in 0.3in 5.6in 0.0in,clip=true, width=40mm}
\epsfig{file=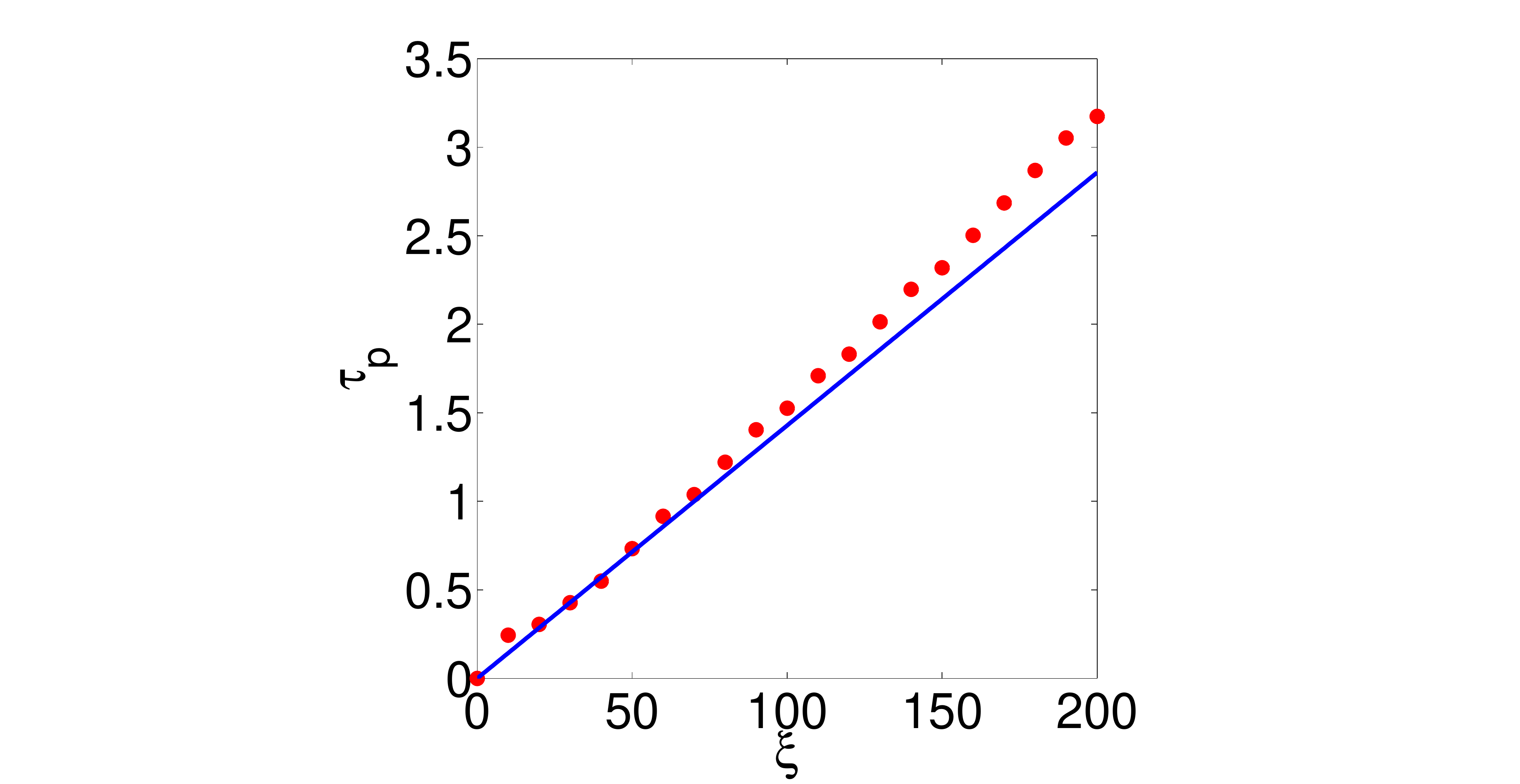,trim=3.3in -0.1in 3.8in 0.0in,clip=true, width=41mm}
~~~~~~~~~~~~~~(a)~~~~~~~~~~~~~~~~~~~~~~~~~~~~~~~~~~(b)
\caption{(Color online) (a) Temporal evolution of ODS under the TOD perturbation.
Input (dotted trace) and output pulse shapes are shown on top. (b) Changes in ODS position with distance as predicted by the variational technique (solid blue line) are compared with numerical data (red circles). The parameters used in the simulations are: $K=0.01,~g_0=0.001,~g_2=0.001$ and $\delta_3=0.1$.}  \label{variational}
\end{center}
\end{figure}

Figure~\ref{variational}(a) shows the evolution of ODS under TOD acting as the sole perturbation using $\delta_3=0.1$. We observe that the ODS nearly preserves its shape with only small variations in the pulse width (mild breathing). The pulse shape at $\xi=200$ is plotted on top in Fig.~\ref{variational}(a), and it shows a small temporal shift from the initial ODS position. We compare this temporal shift (red dots) with the variational prediction (solid line) in  Fig.~\ref{variational}(b). The two agree reasonably well for up to $\xi=100$ with increasing departure for longer distances. This agreement is expected only for relatively low values of $\delta_3$. Indeed, significant distortions of the pulse shape are observed for high values of $\delta_3$. Under weak TOD perturbation, the ODS maintains its overall shape over relatively long distances, and the variational analysis works reasonably well in that situation.

\section{CONCLUSIONS}

By exploiting the standard variational technique, we study the dynamics of a perturbed dissipative soliton excited inside an active semiconductor waveguide. The pulse evolution is governed by an extended GLE containing additional terms that have their origin in higher-order effects such as TOD, self-steepening, Raman scattering, and free-carrier generation. We treat these terms as small perturbations and carry our variational analysis after choosing a dissipative soliton as our ansatz. Being an exact solution of the unperturbed GLE, this chirped soliton maintains its shape inside the active waveguide with slow evolution of its parameters with distance. The variational treatment provides with a set of coupled ordinary differential equations. We have shown that solution of this set of equations predicts quite well how the individual pulse parameters will evolve with distance. we solve the GLE numerically using the split-step Fourier method and show that the variational predictions agree well with full numerical simulations. We also propose simple analytical solutions for the Raman-induced spectra red-shifts and the corresponding temporal shift of the pulse peak. With suitable approximations, our closed-form expressions should prove useful in practice. In summary, our semi-analytical treatment provides significant insights in understanding the complex dynamics of perturbed dissipative solitons.

\section*{Acknowledgements}
This work is supported by SRIC, Indian Institute of Technology, under the project ISIRD. A.S. acknowledges MHRD, India for a research fellowship.


\begin{thebibliography}{99}
\bibitem{Akhmediev}
    N. Akhmediev and A. Ankiewicz, Dissipative Soliton: Lecture Notes in Physics, (Springer, Berlin, 2005).
\bibitem{Bondeson}
    A. Bondeson, M. Lisak and D. Anderson, ``Soliton Perturbations: A variational principle for the soliton parameters," Phys. Scripta $\bf{97}$, pp. 479-485 (1979).
\bibitem{Kaup}
    D. J. Kaup and B. M. Malomed, ``The variational principle for nonlinear waves in dissipative systems,"  Physica D $\bf{87}$, pp. 155-159(1995).
\bibitem{Anderson}
    D. Anderson, ``Variational approach to nonlinear pulse propagation in optical fibers,"  Phys.\ Rev.\ A $\bf{27}$, 3135 (1983).
\bibitem{Cerda}
    S. C. Cerda, S. B. Cavalcanti and J. M. Hickmann, ``A variational approach of nonlinear dissipative pulse propagation," Eur. Phys. J. D $\bf{1}$, pp. 313-316 (1998).
\bibitem{Royjlt}
    S. Roy and S. K. Bhadra, ``Solving soliton perturbation problems by introducing Rayleigh's dissipation function," J. Lightwave Technol. $\bf{26}$, pp. 2301-2322 (2008).
\bibitem{GPAbook1}
    G. P. Agrawal, Nonlinear Fiber Optics, 5th ed. (Academic Press, 2013).
\bibitem{PS}
    N. R. Pereira and L. Stenflo, ``Nonlinear Schroedinger equation including growth and damping," Phys. Fluids $\bf{20}$, pp. 1733-1734 (1977).
\bibitem{Agazzi}
    L. Agazzi, J. D. B. Bradley, M. Dijkstra, F. Ay, G. Roelkens, R. Baets, K. W\"{o}rhoff and M. Pollnau, ``Monolithic integration of erbium-doped amplifiers with silicon-on-insulator waveguides," Opt. Express $\bf{18}$, pp. 27703-27711 (2010).
\bibitem{Dieter}
    D. K. Schroder, R. N. Thomas, and J. C. Swartz, ``Free-carrier absorption in silicon," IEEE J. Solid-St. Circ. $\bf{13}$, pp. 180-187 (1978).
\bibitem{Tomita}
    A. Tomita, ``Free-carrier effect on the refractive index change in quantum-well structures," IEEE J. Quantum Electron. $\bf{30}$, pp. 2798-2802 (1994).
\bibitem{Lin}
    Q. Lin, O. J. Painter and G. P. Agrawal, ``Nonlinear optical phenomena in silicon waveguides: Modeling and applications," Opt. Express $\bf{15}$, pp. 16604-16644 (2007).
\bibitem{Roy-M-B}
    S. Roy, A. Marini and F. Biancalana, ``Self-frequency blueshift of dissipative solitons in silicon-based waveguide," Phys. Rev. A $\bf{87}$, 065803 (2013).
\bibitem{Anderson_Scr}
    D. Anderson, F. Cattani and M. Lisak, ``On The Pereira-Stenflo Solitons," Phys. Scripta. $\bf{T82}$, pp. 32-35 (1999).
\bibitem{GPAbook2}
    G. P. Agrawal, Application of Nonlinear Fiber Optics, 2nd ed. (Academic Press, California, 2005)
\bibitem{Rong}
    H. Rong, A. Liu, R. Nicolaescu, M. Paniccia, O. Cohen and D. Hak, ``Raman gain and nonlinear optical absorption measurement in a low-loss silicon waveguide," Appl. Phys. Lett. $\bf{85}$, pp. 2196-2198 (2004).
\bibitem{Lin-Z-P}
    Q. Lin, J. Zhang, G. Piredda, R. W. Boyd, P. M. Fauchet and G. P. Agrawal, ``Dispersion of silicon nonlinearities in the near infrared region," Appl. Phys. Lett. $\bf{91}$, 021111 (2007).
\bibitem{Dinu}
    M. Dinu, F. Quochi and H. Garcia, ``Third-order nonlinearities in silicon at telecom wavelengths," Appl. Phys. Lett. $\bf{82}$, 2954 (2003).
\bibitem{Desurvire}
    E. Desurvire, Erbium-Doped Fiber Amplifiers (Wiley, 1994).
\bibitem{Hasegawa-K}
    H. Hasegawa and Y. Kodama, Soliton in Optical Communication (Oxford University Press, 1995).
\end{thebibliography}
\end{document}